\documentclass[superscriptaddress, nofootinbib, prd]{revtex4}[12pt]
\usepackage{graphicx}% Include figure files
\graphicspath{{figs/}}
\usepackage {amsmath, amssymb, rotating}
\usepackage{footnote}
\usepackage[usenames,dvipsnames]{color}
\usepackage{mdwlist} %to make compact lists: \begin{itemize*} ... \end{itemize*}
\usepackage{slashed}
\usepackage{verbatim}
\usepackage{mathrsfs}

\newcommand{\Lhot}{\Lambda_\text{hot}}
\begin{document}

\title{Relaxation of the Cosmological Constant}
%\title{A Non-singular Cosmology from Vorticity}
\author{Peter W. Graham}
\affiliation{Stanford 
Institute for Theoretical Physics, Department of Physics, Stanford University, Stanford, CA 94305}

\author{David E. Kaplan}
\affiliation{Department of Physics \& Astronomy, The Johns Hopkins University, Baltimore, MD  21218}

\author{Surjeet Rajendran}
\affiliation{Department of Physics \& Astronomy, The Johns Hopkins University, Baltimore, MD  21218}
\affiliation{Berkeley Center for Theoretical Physics, Department of Physics, University of California, Berkeley, CA 94720}

\begin{abstract}
We present a model that naturally tunes a large positive  cosmological constant to a small cosmological constant. A slowly rolling scalar field decreases the cosmological constant to a small negative value, causing the universe to contract, thus reheating it. An expanding universe with a small positive cosmological constant can be obtained, respectively, by coupling this solution to any model of a cosmological bounce and coupling the scalar field to a sector that undergoes a technically natural phase transition at the meV scale.   A robust prediction of this model is a rolling scalar field today with some coupling to the standard model. This can potentially be experimentally probed in a variety of cosmological and terrestrial experiments, such as probes of the equation of state of dark energy, birefringence in the cosmic microwave background and terrestrial tests of Lorentz violation.
\end{abstract}

\maketitle

\section{Introduction}

The observed accelerated expansion of the Universe is well described by the existence of a small cosmological constant.  However, quantum corrections to this quantity are much larger than the observed value.  One might hope that the mysteries of quantum gravity hold the solution, but dangerous contributions come from very well-known physics at scales where spacetime curvature is weak (for example, finite corrections to vacuum energy from the electron mass).  In this regard, the problem can be seen as one of fine-tuning, where contributions, known and unknown, conspire to cancel to generate the small value detected today.

One approach to this puzzle is the introduction of an exponentially large number of universes, 
%connected in spacetime or other ways,
in which the vacuum energy appears to be a random variable taking on different values in each.  Anthropic selection then determines which universe we are likely to appear in, based on the existence of structure or other arguments, and the assumption that a number of other parameters of our universe (such as the baryon-to-photon ratio, the dark matter abundance and the value of the primordial density fluctuations) are the same over this exponentially large number of universes. 

A more natural solution could come from the dynamical relaxation of the cosmological constant in the early universe via a slowly rolling field in a potential.  Indeed, such a model was attempted by Abbott \cite{Abbott:1984qf} and others (e.g.~\cite{Banks:1984tw}), and a similar model was successfully implemented in a solution to the gauge hierarchy problem \cite{Graham:2015cka}.  Attempts to solve the cosmological constant (CC) problem fell short as they invariably result in an empty universe.  This is a robust problem with relaxation: the CC can only be sensed through gravity, but gravity is universal and thus couples to the total space-time curvature.  One way to solve this empty universe problem is to make the universe undergo a bounce after the relaxation of the CC. This framework also solves many other thorny issues that confront solutions of the CC problem such as the problem of cosmological phase transitions affecting the CC after relaxation \cite{Graham:2017hfr}. Motivated by these considerations, in \cite{Graham:2017hfr}, we discussed ways to obtain a bouncing cosmology but did not address the relaxation of the CC. 

In this paper, we present a dynamical relaxation model for the CC problem. Here, a rolling scalar field takes the universe from a natural, large positive cosmological constant (CC) to a small negative one.  At this point the universe will begin to contract. The contraction increases the energy in the field(s) responsible for the tuning of the CC. At some density, this increased energy, through a small coupling, reheats other matter. The energy density in this matter blue-shifts as the universe continues to contract. We take as an assumption
that this sector can trigger dynamics that causes the scale factor to bounce at short distances, allowing the universe to expand and produce our observed cosmological history.  This bounce could occur through vorticity as in \cite{Graham:2017hfr}, but any other possible bounce model (e.g.~through NEC violating fluids) would work as well.  The relaxation mechanism is independent of the bounce and cosmology that comes after.  The simplest model (section \ref{sec:simple}) naturally tunes a CC scale as large at 10 MeV to a negative CC of scale 1 meV.  We show how a few additional fields (section \ref{sec:epicycles}) and stages of rolling allow one to scan a CC from scales much higher than 1 TeV, and end with a {\it positive} CC of order the current critical density.  These models are experimentally testable (section \ref{sec:signatures}), through astrophysical, cosmological, and laboratory probes.  

This existence proof factorizes the solution of the CC problem into an infra-red (IR) part that accomplishes the sensitive tuning of the CC and an ultra-violet (UV) sector whose purpose is to accomplish a cosmological bounce at high densities. Importantly, the UV dynamics is decoupled from the IR tuning.  This existence proof highlights the importance of  short-distance descriptions of a cosmological bounce, and presents the opportunity to re-imagine the source of the initial perturbations often credited to inflation.  

\section{Simple Model}
\label{sec:simple}

We now show a simple model that naturally tunes the CC from up to  $\sim (10 \:{\rm MeV})^4$ to $\sim -(1 \:{\rm meV})^4$ and subsequently reheats the universe when the universe contracts. In the following section, we will show how one can increase the initial cosmological constant and end on a small positive one.

In this model, a rolling scalar field starts from a point with large vacuum energy  (abiding eternal inflation bounds). As it rolls down, the CC decreases, eventually going through zero. At this point, the universe begins to contract at a parametrically smaller negative CC. The universe contracts to a large energy density, and assuming a cosmological bounce, re-expands until today, while the field value does not evolve significantly, thus keeping the vacuum energy small. Finally, reheating is shown to be trivially accomplished by extracting energy from the rolling condensate and dumping it into a thermal bath through derivative couplings.

Remarkably, all of this can be accomplished with the dynamics of the following example model: 

\begin{equation}
{\cal L} = \frac{1}{2}(\partial\phi)^2 + \frac{1}{4} F'F' +  \bar{\psi}\,(i\, \slash{\!\!\!\!D} - m_\psi)\psi - \frac{1}{2} m_{A'}^2 A'^2 + g^3
\phi - \frac{\phi}{f}F'{\tilde F}',
\end{equation}
Here, $\phi$ is a scalar field with a softly broken shift symmetry, $A'^\mu$ is a massive photon whose gauge field strength is $F'$ and $\psi$ is a charged massive Dirac fermion.  We use a mostly negative metric and we have defined the value $\phi=0$ to be the point of vanishing cosmological constant for convenience. The rolling of $\phi$ decreases the CC and its kinetic energy is eventually converted to the gauge bosons $A'$. The energy in this radiation reheats the universe, producing both the standard model and the degrees of freedom necessary to cause the universe to bounce. 

\subsection{Rolling to -meV$^4$}

\begin{figure}
\includegraphics[scale=0.85]{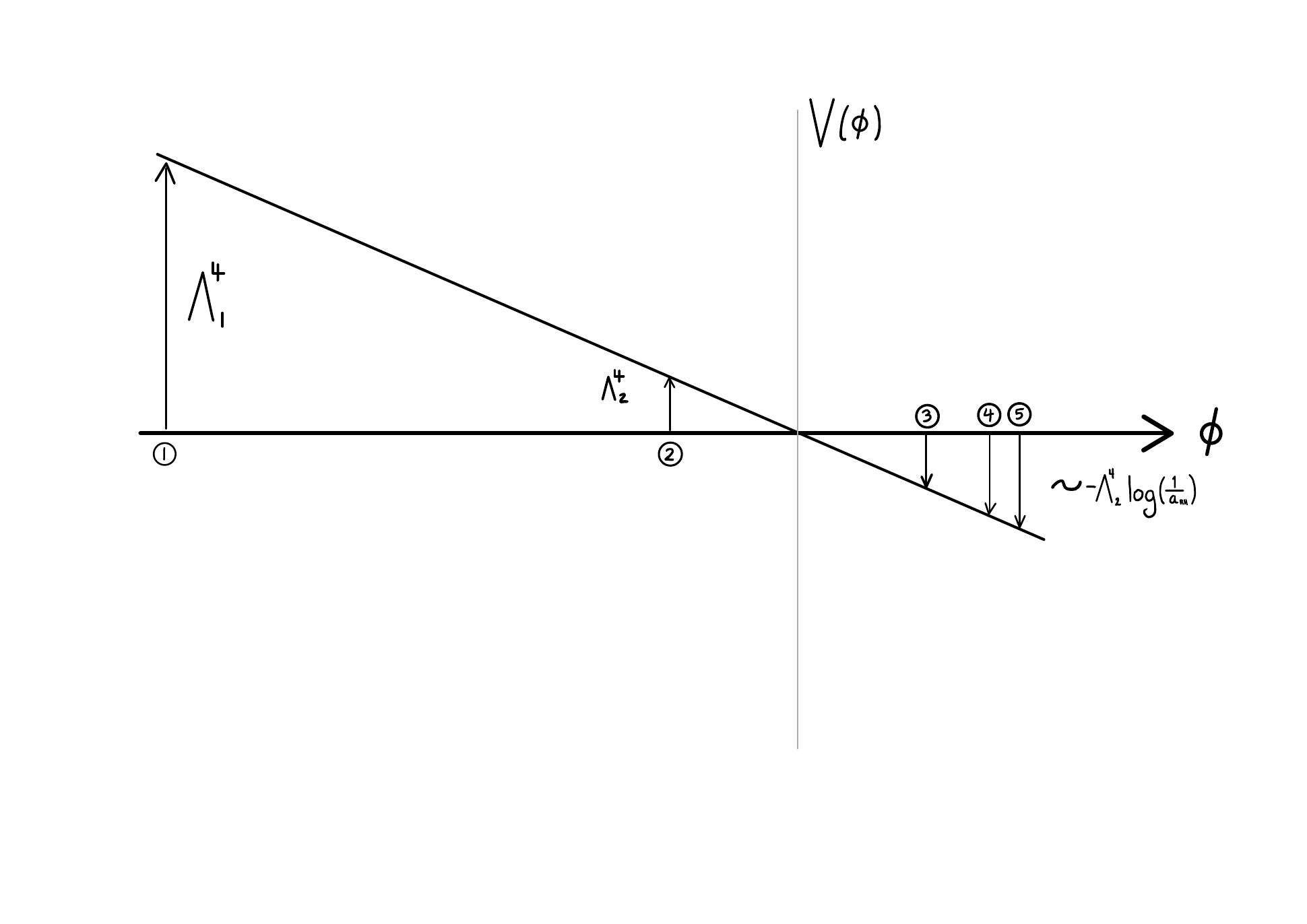}
\caption{\label{roll1} Evolution of $\phi$.  The field starts rolling from point 1 with vacuum energy $\Lambda_1^4$.  At point 2, $\phi$'s kinetic energy equals its potential energy, $\Lambda_2^4$, a distance $\sim M_p$ away from the origin.  At point 3, the Hubble scale passes through zero and the vacuum energy is $\sim-\Lambda_2^4$. The point 4 represents the position of $\phi$ after a period of kinetic-energy dominated contraction, where the vacuum energy as decreased to $\sim -\Lambda_2^4 \log{(1/a_{rh})}$ and $a_{rh}$ is the scale factor at reheating.  At point 5, the reheated universe has expanded until today and $\phi$ has moved a negligible amount from point 4.}
\end{figure}

The dynamics of interest (see Figure \ref{roll1}) start with an initial condition of an expanding universe, and a large negative value for $\phi$, namely $\phi=\phi_1 < 0$ and $|\phi_1|\gg M_{p}$ (the reduced Planck scale).  Here, there is positive vacuum energy,  $\Lambda_1^4 \equiv g^3 \phi_1$, and assuming initially $\dot{\phi}\ll\Lambda_1^2$, $\phi$ slow rolls down its potential in a vacuum dominated universe.  

We would like the universe to evolve to a small cosmological constant and avoid eternal inflation.  This puts the constraint
\begin{equation}
g^3 > \sqrt[]{3/2\pi} H^3 \simeq \sqrt[]{1/2\pi} \Lambda_1^6/(3 M_p^3)
\end{equation}
or a limit on the highest CC that can be relaxed of
\begin{equation}
\label{eqn: eternal constraint}
\Lambda_1 \lesssim \sqrt[]{g M_p}.
\end{equation}

The rolling continues until $\phi$ reaches a value $\phi=\phi_2 \sim - M_p$, where the kinetic energy surpasses the potential energy and $\phi$ is less than a Hubble time away from the origin.  Now, with the kinetic energy increasing, the potential energy becomes negative and the Hubble scale decreases at an increasing rate, as one can clearly see from the Friedmann Equations:
\begin{eqnarray}
H^2 &=& \frac{8\pi}{3}G_{N}\left(\frac{1}{2}\dot{\phi}^2 - g^3 \phi\right) \label{fried1} \\
\dot{H} &=& - 4\pi G_{N} \dot{\phi}^2 \label{fried2}
\end{eqnarray}
The Hubble rate $H$ vanishes in a finite time when a value $\phi\sim M_p$ is reached.  To see this analytically, take Eq. (\ref{fried2}) and integrate from point $\phi_2$ to where Hubble vanishes:
\begin{eqnarray}
0 - H_2 &=& -\int 4\pi G_N \dot{\phi}^2 dt = - 4\pi G_N \int \dot{\phi}d\phi \\
H_2 &>& 4\pi G_N \dot{\phi}_2 \Delta\phi
\end{eqnarray}
where the 2 subscript indicates the values at point $\phi_2$ where kinetic and potential energy are equal.  The inequality comes from the fact that $H$ is monotonically decreasing from (\ref{fried2}) making $\dot{\phi}$ monotonically increasing due to its equation of motion.  This allows us to replace $\dot{\phi}$ in the integral with its minimum (initial) value to generate the inequality.  

Thus, $\phi$ traverses a finite distance (of order $M_p$) in a finite time (as can also be shown numerically). Because $\dot{H}<0$ at this point, $H$ continues to decrease below zero and the universe begins to contract.  The potential energy at this point is $\sim -\Lambda_2^4\equiv - g^3 M_p$.

\subsection{Kinetic Energy during Contraction}

As the universe contracts, the kinetic energy of $\phi$ quickly dominates the potential energy and blue-shifts as $\dot{\phi}^2 = \dot{\phi}_0^2 a^{-6}$, where $a$ is the scale-factor of the Friedmann-Robertson-Walker metric and $\dot{\phi}_0$ is the velocity of $\phi$ at the point where $H=0$ (taking $a=1$ at that point).  Taking the kinetic energy to be dominant, it is simple to compute the distance $\phi$ travels down its potential while the universe contracts:
\begin{eqnarray}
\Delta\phi &=& \int \dot{\phi} \, dt \nonumber\\
&\approx& - \int \dot{\phi}_0 \frac{1}{a^3} \frac{\sqrt{2}}{\sqrt[]{8\pi G_N/3} \,\dot{\phi}_0} a^2 \, da \nonumber\\
&\approx& \sqrt[]{6} M_p \log{(1/a)}
\label{eqn: contraction}
\end{eqnarray}
where we used the first Friedmann equation (\ref{fried1}), the fact that $H<0$, and the definition of the reduced Planck mass. We see that the distance traveled by $\phi$ (and thus the change in the potential energy) is only logarithmically sensitive to the scale factor.  For example, if the universe contracts to a scale where the energy density is $\frac{1}{2} \dot{\phi}^2 = \Lhot^4$, where $a = (\Lambda_{2}/\Lhot)^{2/3}$, then $\Delta\phi = \sqrt[]{6} M_p (2/3)\log{(\Lhot/\Lambda_2)}$, or about $60 M_p$ for $\Lhot=1\; {\rm TeV}$ and $\Lambda_2 = 1 \;{\rm meV}$.  This defines the point 4 in Figure \ref{roll1}.  And so in the time that $\phi$ rolls parametrically not far beyond point 3, the energy density in the universe (kinetic energy in $\phi$) increases all the way up to an essentially arbitrarily high scale $\Lhot^4$.

The final limit on the highest CC that can be naturally scanned down to $\sim {\rm meV}^4$ then arises as follows.  In order to avoid tuning we want the negative CC reached after contraction to be $\sim {\rm meV}^4$ in magnitude.  The value of the CC at point 5 in Figure \ref{roll1} is very close to the value at point 4 (as we will see below).  So we need the value at point 4 $g^3 \phi_4 \lesssim {\rm meV}^4$.  Combining this with equation \eqref{eqn: contraction} and equation \eqref{eqn: eternal constraint} gives 
\begin{equation}
\label{eqn: cutoff limit}
\Lambda_1 \lesssim \left( \frac{3}{2 \sqrt{6}  \log{(\Lhot/\Lambda_2)} } \right)^\frac{1}{6} \, {\rm meV}^\frac{2}{3} \, M_p^\frac{1}{3}
\end{equation}
Or $\Lambda_1 \lesssim 10 \, \text{MeV}$ for $\Lhot=1\; {\rm TeV}$ and $\Lambda_2 = 1 \;{\rm meV}$, with only a very weak dependence on $\Lhot$.  So this model can naturally reduce a CC of 10 MeV to the observed value, reducing fine-tuning by roughly 40 orders of magnitude.

\subsection{Reheating $A'$}

Now we utilize $\phi$'s coupling to the massive vector to convert the kinetic energy of $\phi$ into a thermal bath of $A'_\mu$ and $\psi$.    We describe this process in two stages:

(1st Stage) A population of vectors will be produced when $\dot{\phi}/f > m_{A'}$ due to an instability in the mode equation for the vectors  \cite{Campbell:1992hc, Anber:2009ua, Hook:2016mqo}:
\begin{equation}
\label{eq-sorbo}
\ddot{A'_{\pm}} + (k^2 \pm 4 k\frac{\dot{\phi}}{f} + m_{A'}^2)A'_{\pm} = 0
\end{equation}
where $A'_{\pm}$ are the spatial Fourier transforms of circularly polarized modes of the vector $A'$. The $A'_{-}$ modes with wave numbers $k<(\dot{\phi}/f)$ and $k>(f m_{A'}^2/\dot{\phi})$ will be exponentially growing modes.  Assuming an initial fluctuation of order $A'_{-}\sim k$ in each mode (the minimum set by quantum mechanics), the largest energy density comes from modes $k \sim \frac{\dot{\phi}}{f}$ in the massive vector and so grows up to
\begin{equation}
\label{eq-initialrho}
\rho_{A'}\sim (\dot{\phi}/f)^4 e^{2(\dot{\phi}/f) t_{th}} ,
\end{equation}
where $t_{th}$ is the thermalization time scale.  This initial density sources the thermal destruction of the condensate as we see below.

In order to be sure that the thermal calculations we rely on below (in \eqref{eqn: thermal friction}) are valid, we would like to have the $A'$ thermalize when their energy density is above $m_{A'}^4$ so that the temperature they thermalize at is above $m_{A'}$.  Even if this is violated, the thermal friction may well still stop the rolling of $\phi$ as we want, but it is outside the regime of validity of the thermal field theory calculations that have been done, so we will choose to avoid this region.  As we will see this is easy to do.

We compute a process that will start thermalizing the $A'$.
Assuming the fermion $\psi$'s mass is of order $m_{A'}$, the cross section for scattering $A' A' \to \psi \psi$ is
\begin{equation}
\sigma v \sim \frac{\alpha^2}{m_{A'}^2}
\end{equation}
where $\alpha = \frac{e^2}{4 \pi}$ and $e$ is the charge of $\psi$, and we have assumed that $\frac{\dot{\phi}}{f} \sim m_{A'}$ so the $A'$ produced by the rolling $\phi$ are semi-relativistic.  We want the vector to thermalize only after the energy density in $A'$ has reached $m_{A'}^4$.  The number density of $A'$ at this point is $n \sim m_{A'}^3$.  Then the scattering rate at this point is
\begin{equation}
\Gamma_{A' A' \to \psi \psi} \sim n \sigma v \sim \alpha^2 m_{A'}
\end{equation}
We need this to take longer than the time it takes $\phi$ to produce this energy density which is, using (\ref{eq-initialrho}), $t_{th} \sim 2(f/\dot{\phi})\log{(m_{A'} f/\dot{\phi})}$.  Again, assuming $\frac{\dot{\phi}}{f} \sim m_{A'}$, this simply requires that $\alpha<1/\sqrt{2}$.  Satisfying this bound allows the energy density in the vectors to grow to at least $m_{A'}^4$ before thermalization begins.

(2nd Stage) After there are some $A'$ particles around, they will produce $\psi$ particles, and they can then thermalize rapidly.  This then leads rapidly to a thermal bath which will then cause thermal friction of $\phi$ (see e.g.~\cite{Laine:2016hma}).  We will take the thermal friction coefficient to be:
\begin{equation}
\label{eqn: thermal friction}
\Gamma \sim 256 \pi^2 \alpha^3 \frac{T^3}{f^2}
\end{equation}
where $T$ is the temperature of the thermal bath, and the coupling $\alpha$ is the renormalized coupling at $T$.  This is the result for the analogous friction in the case of pure Yang-Mills, and the numerical factor is a result of normalization difference with \cite{Laine:2016hma}.  There may be numerical differences between the computed Yang-Mills case and the yet to be computed Abelian case with charged fermions, but it is unlikely to be parametrically different (at least with respect to $f$ and $T$ dependence), and thus we will use this value as a rough estimate of the friction.\footnote{One can also build a model where the sector being heated is non-Abelian and the mass scale $m_{A'}$ corresponds to the confinement scale of the strong group.  In order to avoid the generation of larger barriers in the $\phi$ potential, on can add an additional massless quark.  And also, this model is close enough to that studied in \cite{Laine:2016hma} to suggest the damping rate $\Gamma$ is parametrically the same.}

This coefficient will damp and ultimately suppress the $\phi$ rolling such that its energy density is negligible relative to that of the radiation.  We want to see how long this process takes.  Initially while this damping is happening, the bath is at some temperature $T$ which is less than the kinetic energy in $\phi$.  Then the rate at which energy density is being taken out of the $\phi$ rolling and put into the thermal bath is
\begin{equation}
\frac{d \rho}{dt} \sim 256 \pi^2 \alpha^3 \frac{T^3}{f^2} \dot{\phi}^2
\end{equation}
Setting the energy density in the thermal bath to $\rho \sim T^4$ we find
\begin{equation}
\frac{d T}{dt} \sim 64 \pi^2 \alpha^3 \frac{\dot{\phi}^2}{f^2} 
\end{equation}
This means that it takes the longest time to get it up to the highest temperature (the time is dominated by the UV).
We can see how long it takes to remove an O(1) fraction of the kinetic energy in $\phi$ by setting $\dot{\phi}$ to be a constant equal to its initial value when we want to reheat: $\dot{\phi} \sim T_\text{reheat}^2$.  Then we see how long it takes $T$ to get up to this value, call this time $\Delta t_\text{reheat}$.  We can see that
\begin{equation}
\Delta t_\text{reheat} \sim \frac{f^2}{64 \pi^2 \alpha^3 T_\text{reheat}^3}
\end{equation}
Requiring that this happen within a Hubble time then means that we set $\Delta t_\text{reheat} \lesssim \frac{1}{H} \sim \frac{M_p}{T_\text{reheat}^2}$.  This then leads to the requirement that
\begin{equation}
\label{eqn: reheat constraint}
f^2 \lesssim 64 \pi^2 \alpha^3 T_\text{reheat} M_p .
\end{equation}
Plugging in $\dot{\phi} \sim T_\text{reheat}^2 \sim f m_{A'}$, we also have the constraint $f \lesssim 16 \pi^{4/3} (m_{A'} M_p^2)^{1/3}$.  From equation \eqref{eqn: reheat constraint} we can see that $f$ has a wide range of possible values even for a fairly low reheating temperature.  And $f$ can even be all the way up near $M_p$ if we take a high reheating temperature.

\subsection{Bouncing}

The contracting universe needs to bounce (evolve to an expanding universe) so that it can re-expand and reproduce our cosmic history. The dynamics responsible for the bounce can be decoupled from the tuning of the cosmological constant. This is easily accomplished - after all the kinetic energy of $\phi$ is dumped by thermal friction into $A'$, the $A'$ can reheat the degrees of freedom responsible for the bounce through weak couplings. Around the time of the bounce, these degrees of freedom must effectively violate the null energy condition or be able to trigger vorticity in extra dimensions as in \cite{Graham:2017hfr}. In order to bounce, these degrees of freedom need to blue-shift faster than the other matter content in the universe so that they are relevant at the short distances where the bounce occurs. Further, when the universe re-expands after the bounce, this matter must return to its original state so that the tuning of the CC is not affected. This can likely be guaranteed if the behavior of this sector is determined by thermodynamics (such as a temperature), wherein the re-expansion of the universe would cool this sector, returning it to its original state. However, we leave an explicit model of the way to trigger the bounce starting with our CC relaxation model for future work.

\begin{comment}
These degrees of freedom must effectively violate the null energy condition. In order to bounce, these degrees of freedom need to blue-shift faster than the other matter content in the universe so that they are relevant at the short distances where the bounce occurs. 

As a toy example, imagine the universe, while contracting, has three components of energy density -- the radiation from reheating, $\rho_r$, and much larger and (almost) mutually equal positive and negative contributions to vacuum energy $\rho_\pm$.  When $\rho_r$ blue-shifts to the point where $\rho_r \sim |\rho_-|$, a coupling between the two could cause a transition such that $\rho_$ begins to blue-shift with an equation of state $w_- > 1/3$ (or $w=1/3$ if $\rho_r<\rho_-$ at the moment of transition).  The negative energy would grow faster than the positive energy and eventually the Hubble parameter $H$ would pass through zero -- a bounce.  Then as the universe expands, the transition should reverse and $\rho_-$ must settle back down to its original value (so as not to mess up the tuning of the CC).  A similar scenario could use $\rho_+$, forcing it to transition into energy density with an equation of state $w_+ < -1$ causing it to {\it red shift} as the universe contracts.  Both scenarios clearly require new matter that violates the null energy condition. 
\end{comment}

Before the bounce $\phi$ rolls only a short distance as seen above.  And all the initial kinetic energy of $\phi$ from contraction is dumped into the $(A'_\mu,\psi)$ sector by thermal friction, so $\phi$ is then moving very slowly.  Once the universe bounces and re-expands, it is dominated by radiation, specifically in the  $(A'_\mu,\psi)$ sector, and the kinetic energy in $\phi$ is never above its terminal velocity value $\dot{\phi}^2\sim g^6/\Gamma^2$, which (as can be easily shown) keeps $\phi$ from rolling a significant amount.  At any time, even if thermal friction becomes smaller than Hubble friction, $\phi$ will not roll more than $\Delta \phi \sim (\Lambda_2/T)^4 M_p$ in a Hubble time.  So $\phi$ does not roll significantly during the entire contaction, bounce, and subsequent expansion of the universe.  Thus the potential energy of $\phi$ is not changed significantly and so the dynamical relaxation solution for the CC is not spoiled.

\subsection{Reheating the Standard Model}

The last step would be to reheat the rest of the universe (namely the Standard Model sector). This can be accomplished by coupling the massive vector to the normal matter through mixing with the photon or through higher dimensional operators, and allowing the Standard Model to thermalize at some point when the temperature is higher that the scale of big bang nucleosynthesis.

A kinetic mixing with the hypercharge gauge boson, $\epsilon F'_{\mu\nu}F_Y^{\mu\nu}$ would allow the vector $A'$ decay into standard model particles with a rate $\Gamma_{decay} \sim \alpha_Y \epsilon^2 m_{A'}$.  Equating this with the Hubble scale, this gives the temperature of the Universe at the time of decay: $T_d \sim \alpha_Y^{1/2}\epsilon \sqrt{m_{A'}M_p}$.  On the other hand, the mixing of the vectors produces an effective coupling to $\phi$ of the form  $\epsilon^2 (\phi/f) F^Y_{\mu\nu}\tilde{F}_Y^{\mu\nu}$.  If we require -- though it may not be necessary -- the rate of this instability (from the analogous version of equation (\ref{eq-sorbo}) for photons), to be less than Hubble, $\dot{\phi}\epsilon^2/f < H$, then none of the dynamics described above change.  This constraint is most sensitive at the lowest values of Hubble, where $\dot{\phi} \sim g^3/H_2\sim H_2 M_p$, thus requiring $f> \epsilon^2 M_p$.  One can show that these constraints are trivial to satisfy.

\section{Epicycles}
\label{sec:epicycles}

The model presented above  naturally takes a large cosmological constant and relaxes it to a parametrically smaller (albeit negative) one, converts the energy from this sector to a hot standard model, and (after a bounce) produces normal big bang cosmology with a tiny cosmological constant.  In this section, we will show how a few additional degrees of freedom will allow: (a) dynamics that produce a small {\it positive} cosmological constant and (b) natural relaxation from much larger cosmological constants, while maintaining technical naturalness. 

\subsection{Positive CC}

Suppose the CC has been reduced to  $\sim - \text{meV}^4$, with the energy in the rolling field dumped into other forms of matter. At this stage, the universe starts crunching and the energy density in these matter fields will blue-shift. This energy can be used to trigger  a technically natural phase transition at the $\sim \text{meV}^4$ scale, resulting in an addition to the vacuum energy and the CC changing from $\sim - \text{meV}^4$ to $\sim + \text{meV}^4$.  This transition is not fine-tuned so long as the CC is changed to a positive value of roughly the same size as, or greater than, the small negative value it had after relaxation.  Once the universe has already started to crunch, changing the CC by $\sim + \text{meV}^4$ does not change the dynamics of the universe as its energy density is dominated by the rapidly blue-shifting matter or radiation density. Thus, the rest of the cosmic evolution necessary to implement our framework such as the bounce and the subsequent re-expansion of the universe are unaffected by the transition necessary to achieve a positive CC.  There are likely many ways to accomplish this goal, we present one such example in  Appendix \ref{sec:axioncc}. 

\subsection{Larger Cutoff} 

The principal difficulty in achieving a larger cutoff in the model presented in section \ref{sec:simple} is that the slope of $\phi$ needs to be sufficiently large when the vacuum energy is big in order to avoid eternal inflation. This large slope induces a kinetic energy for $\phi$ that  causes it to roll well beyond $\sim \text{meV}^4$. To achieve a larger cutoff, we can introduce an additional rolling scalar field ($\Phi$) that has a steeper slope. We start the universe with both $\phi$ and $\Phi$ rolling. When the CC is large, the rolling of $\Phi$ provides the clock necessary to avoid eternal inflation. As the CC approaches $\sim \left(10 \text{ MeV}\right)^4$, we need to create barriers that stop the rolling of $\Phi$. Once $\Phi$ is stopped at $\sim \left(10 \text{ MeV}\right)^4$, the rolling of $\phi$ will further relax the CC down to $\sim - \text{ meV}^4$. 

How can we naturally trigger barriers for $\Phi$? The key idea is to observe that when the CC is large, the large Hubble friction results in a low terminal velocity for $\Phi$. As the CC drops, Hubble friction decreases, resulting in a larger terminal velocity.  We use this to trigger the barriers. The increased velocity of $\Phi$ can trigger instabilities in gauge fields to which $\Phi$ is derivatively coupled, such as in the models discussed in section \ref{sec:simple}. The energy released in this process can be used to raise barriers for $\Phi$. There are many ways to accomplish this goal - we present a proof of concept model in Appendix \ref{sec:highcutoff}. Note, this initial stage of relaxation does not require a bounce since the relaxation ends at a relatively high, positive value of the CC. Models of this kind could potentially also be used to simply relax the value of the CC in inflationary relaxion models where the CC could be reduced from the cut-off to the weak scale.   

\section{Potential Signatures}
\label{sec:signatures}

There are four generic elements of our construction: a rolling scalar field $\phi$ that cancels the bare cosmological constant, a bounce in our immediate past to reheat the universe, a phase transition that should occur at scales $\sim$ meV in order to push the cosmological constant to slightly positive values after it becomes negative and strong dynamics at various scales (for example, $\sim$ 10 MeV) that enable the cutoff of the theory to be above the TeV scale. Each of these elements can be separately tested. 

Any dynamical relaxation model requires a field that scans the CC.  So a relatively model-independent signature of this framework is that
the kinetic energy of the rolling field $\phi$ gives rise to a non trivial equation of state for dark energy. Current bounds on the equation of state of dark energy imply that the velocity $\dot{\phi}$ of the field is $\lessapprox \, 0.1 \, \text{meV}^2$ \cite{Aghanim:2018eyx}. In the simplest of our models, we expect $\dot{\phi} \sim g^3 M_{pl}/\text{meV}^2$ with $g^3 \lessapprox \text{meV}^4/M_{pl}$. Thus, current and near future probes of the equation of state of dark energy are constraining the simplest models that can solve the cosmological constant problem. In addition to cosmological probes, the kinetic energy $\dot{\phi}$ can also be probed in laboratory experiments if $\phi$ has couplings to the standard model. Of course, this coupling is necessary at some level since the kinetic energy of $\phi$ has to reheat the universe just before the bounce, ultimately resulting in our existence. Radiative stability of $\phi$ and efficient reheating implies that $\phi$ must couple derivatively to the standard model, much like an axion. There are two leading interactions that can be experimentally probed: the coupling of $\phi$ to electromagnetism and nucleon/electron spins via the operators  $\frac{\phi}{f_a} F \tilde{F}$ and  $\frac{\partial_{\mu} \phi}{f_a}\bar{\Psi} \gamma^{\mu} \gamma_5 \Psi$ respectively. The electromagnetic coupling is already constrained -- if $\dot{\phi}\sim 0.1\: {\rm meV}^2$, current constraints on $B$ modes in the CMB require $f_a \gtrapprox M_{pl}$ as this coupling will cause the polarization of CMB photons to rotate as they propagate through the evolving dark energy \cite{Akrami:2018odb}. Interactions with nucleon/electron spins can potentially be probed through tests of Lorentz symmetry since the evolving dark energy provides a cosmic background that is being searched for in these experiments \cite{Pospelov:2004fj}. These signatures are relatively model-independent signatures of a dynamical relaxation solution to the CC problem.

In our model, a cosmic bounce is required.  This could be detected through a cosmological background of stochastic gravitational waves. The hubble scale during a bounce is not constant - thus, the gravitational wave spectrum would exhibit a sharp feature corresponding to the minimum of the bounce, unlike inflationary cosmology that produces a nearly scale invariant spectrum. The detection of stochastic gravitational waves at different frequency bands would enable experimental discrimination between these two possible cosmological scenarios in our immediate past. 

One of the simplest ways to obtain a slightly positive cosmological constant after the rolling of $\phi$ makes it slightly negative is to reheat a hidden sector that undergoes a phase transition at the $\sim$ meV scale. This suggests that the universe could contain a hidden sector of dark radiation around the meV scale, with a phase transition likely to occur in this sector. Such a transition would also indicate an evolving equation of state of dark energy. Moreover, it would also be interesting to directly search for dark radiation in laboratory experiments, building on the work that has occurred in recent years on searching for ultra-light dark matter. 

Finally, we expect the existence of confining sectors at scales such as $\sim$ MeV in order to push the cutoff of the theory to scales above $\sim$ TeV. It would be interesting to develop techniques to search for such confining sectors - for example, this sector might contain degrees of freedom such as glueballs which can interact with the standard model. A generic operator analysis suggests that these interactions are suppressed. However, since these particles are light, they could conceivably be probed in high statistics intensity frontier experiments \cite{Benato:2018ijc}. 

\section{Discussion and Conclusions}

We have shown a technically natural way to solve the CC problem. Our framework takes a large positive CC and reduces it to a small negative CC through dynamical relaxation. This causes the universe to crunch. At the same time, the relaxation process also naturally dumps energy into a new sector. The energy densities in this sector blue-shifts during contraction thus reheating the universe to high temperatures.  By using the energy in the new sector, we are also able to naturally push the CC to positive values after the universe begins to crunch. This new sector has to ultimately be responsible for instigating a cosmological bounce so that the universe can re-expand, giving rise to the present day universe that had a hot big bang but with a small CC, though we do not model the bounce here. This framework overcomes all the obstacles faced by dynamical relaxation methods to solve the CC problem such as the empty universe problem in Abbott's model, Weinberg's ``no-go'' theorem \cite{Weinberg:1988cp} and the problem of standard model phase transitions. Many aspects of this framework lead to testable consequences, some of which require the development of new experimental probes to target these specific signatures. 

An important fact about this construction is that the  UV dynamics of the bounce are decoupled from the IR relaxation process. Moreover, the CC itself does not change significantly during the bouncing phase. There is thus considerable freedom to attach the IR relaxation phase to any UV dynamics permitting a cosmological bounce \cite{Steinhardt:2001vw, Creminelli:2007aq, Battefeld:2014uga, Creminelli:2006xe,Rubakov:2006pn,Creminelli:2010ba, Rubakov:2014jja, Balasubramanian:2014jaa, Alberte:2016izw, Graham:2017hfr}. To be considered a complete solution to the CC problem, we need to identify the specific mechanism that would allow the hot matter in the crunching universe to trigger a bounce. This requires a better understanding of the matter sources necessary to create a bounce. 

To reproduce observational facts about our universe, it is important to identify mechanisms that would give rise to the scale invariant spectrum of perturbations that have been observed in the CMB. Since our model is largely just inflation, but with reheating accomplished through a bounce, there are elements of scale invariance built into the mechanism. For example, the rolling scalar field will have nearly scale invariant fluctuations until the CC goes through zero. It would be interesting to see if these fluctuations could seed the observed spectrum of perturbations.  At the very least, a period of inflation could follow the bounce.

Another direction worthy of exploration is to see if the relaxion paradigm that solves the hierarchy problem can be successfully incorporated into this CC relaxation mechanism. Cosmological relaxation appears to be the only dynamical mechanism that has the potential to solve the naturalness problems associated with both the CC and the weak scale. These phenomena find a natural home in a universe that is much older than conventionally assumed --- something that is observationally possible and theoretically interesting.

\acknowledgements
  We thank Peter Arnold, Savas Dimopoulos, Michael Fedderke, Nemanja Kaloper,  Mikko Laine,  Guy Moore  and  Raman Sundrum for discussions. SR was supported in part by the NSF under grants PHY-1638509 and PHY-1507160, and the Simons Foundation Award 378243. PWG acknowledges the support of NSF grant PHY-1720397, DOE Early Career Award DE-SC0012012. DEK acknowledges the support of NSF grant PHY-1214000.  This work was supported in part by Heising-Simons Foundation grants  2015-037,  2015-038, and 2018-0765, DOE HEP QuantISED award \#100495, and the Gordon and Betty Moore Foundation Grant GBMF7946.

\appendix

\section{Axion Model for Positive CC}
\label{sec:axioncc}

As discussed in the main article, a simple mechanism for generating a positive CC is by allowing the thermal bath to generate a phase transition to a vacuum with a higher vacuum energy (by an amount meV$^4$).  Here we present an explicit model, though many others are possible.

\begin{figure}
\includegraphics[scale=1.0]{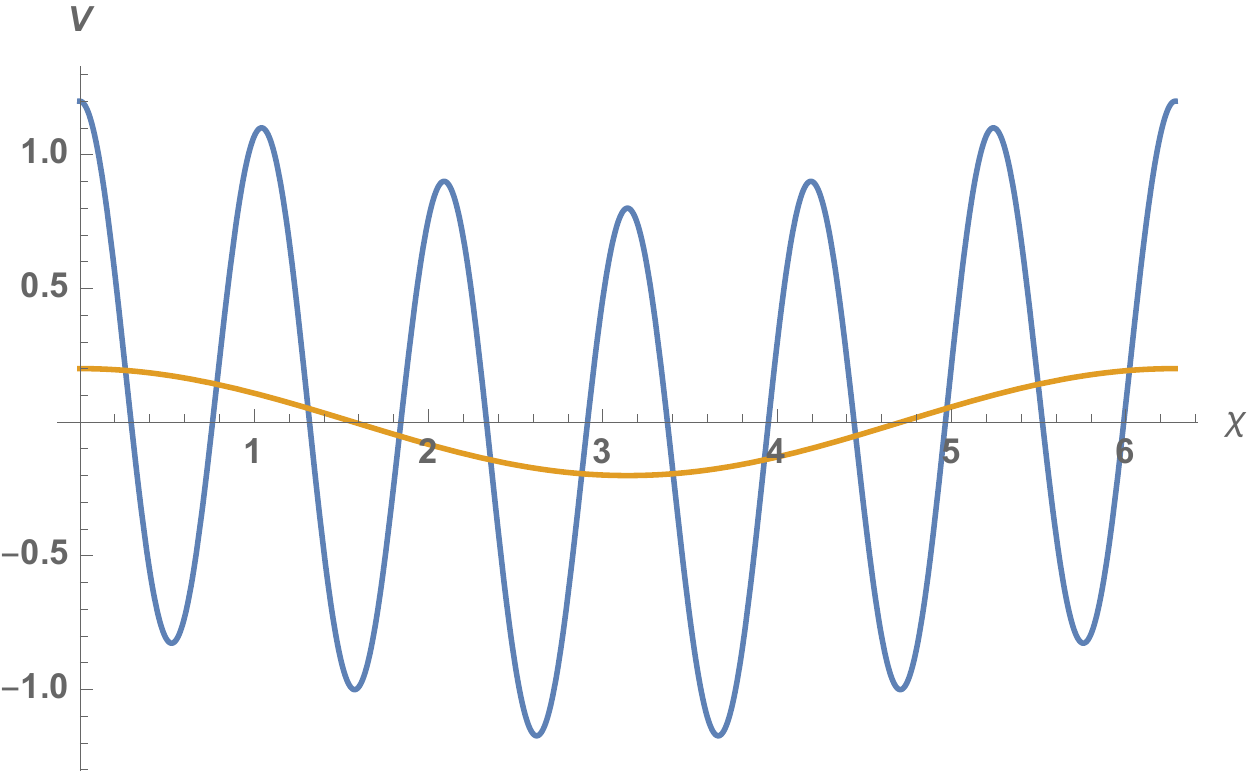}
\caption{\label{fig-positive CC} A sketch of the potential in equation \eqref{eqn: positive CC}.  The blue line is the potential at low temperatures, the orange line is the potential at high temperatures where only the second term in equation \eqref{eqn: positive CC} is on.  This axion begins in some random minimum in its potential and then ends up in a different one.  The vacuum energies of the minima are split by $ {\rm meV}^4$ so this can easily raise the negative CC left after relaxation to a positive CC.}
\end{figure}

Take the following low-energy potential for an axion-like field, $\chi$:
\begin{equation}
\label{eqn: positive CC}
%{\cal L} = \frac{1}{2}(\partial\chi)^2  
V(\chi) =  \Lambda^4 \cos{\frac{n \chi}{f}}  - \tilde{\Lambda}^4 \cos{\frac{\chi}{f}}
\end{equation}
where $n$ is any small integer bigger than 2 and $\lambda > {\rm meV}$ and $\tilde{\Lambda} \sim {\rm meV}$.  This potential is shown in Figure \ref{fig-positive CC}.  We take this potential to be periodic with period $\sim f$.  Take $\chi$ to be in some random minimum after the CC relaxation discussed above. 
During the crunching universe, it is possible for the $\Lambda$ sector to thermalize while the $\tilde\Lambda$ sector does not.  This can happen if the $\tilde\Lambda$ confinement scale is much higher, but the sector has a small quark mass allowing $\tilde{\Lambda}\ll\Lambda$.  During the contracting phase, once the temperature of the $\Lambda$ sector rises  above its confinement scale, the barriers would disappear.  Then $\chi$ will begin to roll toward the minimum of the $\tilde{\Lambda}$ potential.  But then this velocity will rapidly blue-shift because of the contraction.  It is easy to check that there is a large parameter space where this large velocity will cause $\chi$ to go around the entire $\sim f$ period of the periodic potential many times.  As the universe cools, it will ultimately end up in  other random mimimum, in general different from the one it started in.  And there is thus an $\mathcal{O}(1)$ chance it will be a higher minimum with a net positive CC.  This method is not tuned so long as the original relaxation mechanism tuned the CC down to a negative value of $\sim {\rm meV}$ or lower.  Then the amount added to the CC by this axion field is what determines the CC today.

\section{Higher Cutoff}
\label{sec:highcutoff}
In this section, we show a simple, proof of principle, model with one extra rolling field (beyond the one that was already used in the model of Section \ref{sec:simple}) that allows us to push the cutoff well above $\sim$ 10 MeV. More generally, this setup describes how the CC can be relaxed from $\sim \Lambda_{a}^4$ to a lower, positive value $\sim \Lambda_{d}^4$. This stage can precede the model presented in section \ref{sec:simple}, thus instrumenting the full tuning.  In fact it could also replace the model of Section \ref{sec:simple} as the lowest stage of relaxation, except that since it leaves a positive CC the universe does not naturally start crunching and heat up, so there would have to be some other way to avoid the `empty universe' problem.

The model of this stage is also one of a rolling field tuning the CC.  During the rolling, however, the field is coupled to a non-Abelian group (which can generate barriers) plus an Abelian group (to generate additional friction):
\begin{equation}
{\cal L} = \frac{1}{2}(\partial\Phi)^2 + \frac{1}{4} G' G' + \frac{1}{4}F' F' + \kappa^3 \Phi - \frac{\Phi}{f_{th}} G' \tilde{G}' - \frac{\Phi}{f_A} F' \tilde{F}',
\label{eq-stage2}
\end{equation}
where $G'$ and $F'$ are the field strengths of the non-Abelian and Abelian groups (with indices suppressed), and the remaining parameters are couplings.  During the rolling, the coupling of the rolling field, $\Phi$, to non-Abelian gauge bosons produces a non-trivial background temperature for the non-Abelian group (akin to what happens in warm inflation \cite{Rangarajan:2018tte}).  The temperature remains high enough for a long period during which the instanton-generated potential barriers do not form.  Once the Hubble scale becomes low enough, the Abelian group begins to extract energy from the rolling during which the background temperature decreases.  When the temperature become low enough, the barriers form and stop the rolling of $\Phi$, fixing the CC to a value parametrically smaller than its initial value.  During this entire period -- as we will show at the end of this sub-section -- the model of the previous section does not evolve much until Hubble becomes small enough that it can begin to roll consequently.

\begin{figure}
\includegraphics[scale=1.0]{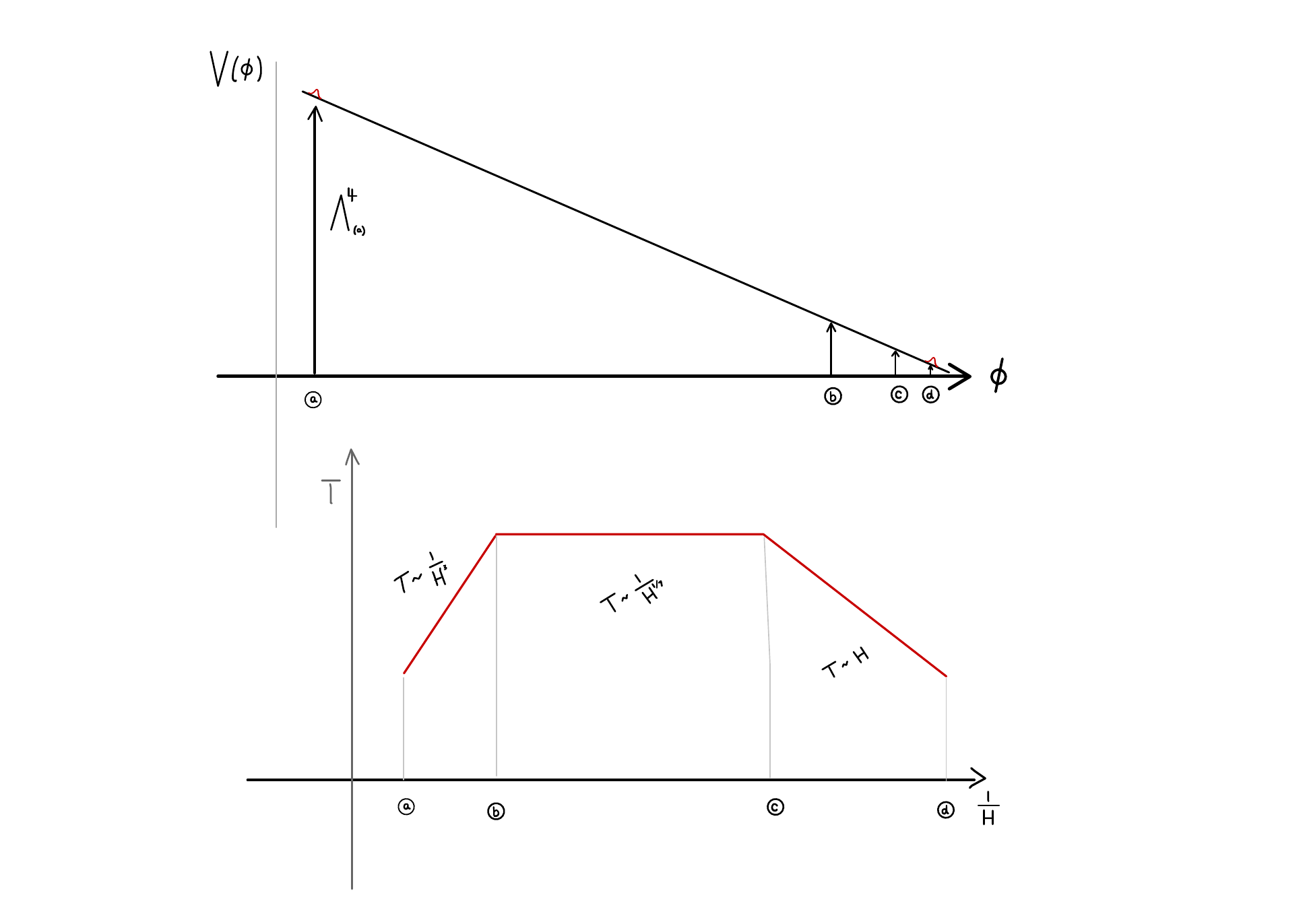}
\caption{\label{fig-stage2} Evolution of $\Phi$.   In the upper figure, the field rolls from (a) where Hubble friction dominates and the background temperature grows quickly, to (b) where thermal friction dominates and the temperature grow very slowly, to (c) where $\Phi$'s rolling is dominantly suppressed by coherent production of Abelian fields and the temperature drops.  At point (d), the temperature drops to the confinement scale of the non-Abelian group and barriers form.}
\end{figure}

Thus, while the model in (\ref{eq-stage2}) is quite simple, the dynamics associated with rolling are quite non-trivial, and we describe them in the following subsections chronologically for a rolling field.  A summary of the motion is presented in Figure \ref{fig-stage2}.

\subsection{Early Friction:  Hubble}

The field $\Phi$ is taken to be slowly rolling (as an initial condition) at point (a) with vacuum energy $\Lambda_{(a)}^4$ and Hubble scale $H_{(a)}\sim \Lambda_{(a)}^2/M_p$.  In addition, there is assumed to be a small background temperature in the non-Abelian group's degrees of freedom, a temperature much smaller than $\Lambda_{(a)}$, but larger than the strong group's confinement scale, which we will define as $\Lambda_{th}$.  The latter will be the scale of the non-perturbative, low-energy potential for $\Phi$.

During to the rolling, the background temperature is maintained in the non-Abelian sector due to thermal friction $\Gamma(T)$ as described in the last section via Eq. (\ref{eqn: thermal friction}).  One can compute an approximate steady-state temperature during this rolling period by requiring the red-shifting/cooling of the plasma bath is counteracted by the heating from the condensate:
\begin{equation}
0 \simeq \frac{d T^4}{dt} \approx -4 H T^4 + \left(256 \pi^2 \alpha'(T)^3\frac{T^3}{f_{th}^2}\right) \frac{1}{2} \dot{\Phi}^2\label{eq-steadyT}
\end{equation}
where in front of $T^4$ there is really a factor that includes the number of degrees of freedom in the non-Abelian sector which we take to be $\sim 1$.
When Hubble friction dominates, we can take $\dot{\Phi}\simeq (\kappa^3/H)$, its terminal velocity during slow roll.  Thus, for temperatures well above the confinement scale, and taking a nominal value of the coupling $\alpha'$  (and ignoring its temperature dependence), one finds a quasi-steady-state temperature of 
\begin{equation}T \simeq  2 \kappa^6/(9 H^3 \bar{f}^2),\label{eq:steadyI}\end{equation}
where we defined $\bar{f}\equiv f_{th}/(16\pi \alpha'^{3/2})$.

\subsection{Late Friction: Thermal}

As $\Phi$ rolls, the Hubble scale will slowly drop and the temperature will rise, eventually reaching a point when the dominant friction is due to the thermal bath, {\it i.e.}, $\Gamma(T)\simeq T^3/\bar{f}^2 \sim H$.  This again produces a terminal velocity for $\Phi$, namely $\dot{\Phi}\simeq \kappa^3/\Gamma\simeq \kappa^3 \bar{f}^2/(T^3)$ , and a temperature, using (\ref{eq-steadyT}), of
\begin{equation}
T\simeq \left( \frac{\kappa^6 \bar{f}^2}{8 H}\right)^{\frac{1}{7}} .
\label{eq:steadyII}
\end{equation}
As Hubble slowly decreases, the temperature stays nearly constant.  Without the Abelian sector, the temperature would eventually dominate Hubble and the universe would become radiation dominated (as in warm inflation \cite{Anber:2009ua}), and $\Phi$ would roll to negative values of the cosmological constant.  We instead would like the temperature to drop before this happens.

\subsection{Mode Instability:  Cooling}

The coupling of $\Phi$ to the Abelian sector produces an instability in some modes of the gauge fields.  An effective negative mass term appears in the equation of motion for the $A'_+$ modes with $k<\dot{\Phi}/f_A$ as in equation (\ref{eq-sorbo}), but with $m_{A'}$ set to zero.  Following the analysis of \cite{Anber:2009ua}, one can show that a quasi-steady-state is reached when a rolling field is coupled in this way to an Abelian group and vacuum energy dominates the energy density of the universe.  In this regime, 
\begin{equation}
\dot{\Phi} \simeq \xi f_A H\label{eq-sorboSS}
\end{equation}
where 
\begin{equation}
\xi = \frac{1}{2\pi} \log{\left(\frac{9 \cdot 2^{21}\pi^2}{7! \,  \alpha_A}\frac{M_p^4 f_A \kappa^3}{ V(\Phi)^2}  \right)} \sim  10 - 100
\end{equation}
where $\alpha_A$ is the fine-structure constant for the Abelian group.  The velocity in (\ref{eq-sorboSS})  becomes the terminal velocity when this instability becomes the dominant energy-loss mechanism, which is equivalent to when (\ref{eq-sorboSS}) is smaller than the terminal velocity due to thermal friction, $\kappa^3/\Gamma$ which happens when:
\begin{equation}
H< \frac{\kappa^3 \bar{f}^2}{ \bar{f}_A T^3}
\end{equation}
where for simplicity, we define $\bar{f}_A\equiv \xi f_A$.  With this velocity, thermal friction still extracts energy from the rolling generating a quasi-steady state temperature:
\begin{equation}
T\simeq \frac{\bar{f}_A^2 H}{8 \bar{f}^2}
\label{eq:steadyIII}
\end{equation}
which thus decreases as $\Phi$ rolls.  The rolling thus eventually stops as the temperature drops to the confinement scale of the non-Abelian group and barriers in the $\Phi$ potential (from instanton effects) begin to form.  This must occur while vacuum energy is still dominating, both over the temperature bath and over the energy density in the Abelian fields, which is estimated to be \cite{Anber:2009ua} $\kappa^3 \bar{f}_A $.

\subsection{Constraints on the Initial Cosmological Constant}

Now we can use the above behavior to put constraints on parameters, including the initial value of the CC.  The constraints are as follows:
\begin{align*}
    \Lambda_{th} &\lesssim T_{(a)} && \text{Temp is higher than confinement scale at start.} \\
    \Lambda_{th} &\sim T_{(d)} && \text{Temp lowers to confinement scale at end.}\\
    \Lambda_{th}^4 &> \kappa^3 f_{th} && \text{Barrier's slope beats underlying slope.}\\
    \Lambda_{(d)}^4 &> \kappa^3 \bar{f}_A && \text{CC dominates over Abelian mode growth.}
\end{align*}
Additional constraints, such as the requirement that the final vacuum energy is greater than the energy density in the thermal bath and that thermal fluctuations do not lead to eternal inflation at any time during the scan, can be shown to be weaker than those above for the ranges of parameters here.

Combining these constraints with the steady-state temperatures at the beginning and end of $\Phi$'s roll (Equations \eqref{eq:steadyI} and \eqref{eq:steadyIII}, respectively), one can derive the following bound:
\begin{equation}
\Lambda_{(a)} < \frac{\Lambda_{(d)}^{35/27} M_{p}^{10/27}}{\bar{f}^{2/3}}
\end{equation}
where we have taken $\alpha'={\cal O}(1)$.  A strict constraint on $\Lambda_{(a)}$ could come from requiring $f_{th}=16\pi \alpha'^{3/2}\bar{f}$ to be greater than $\Lambda_{(a)}$ (the CC at top), producing:
\begin{equation}
\Lambda_{(a)} < \Lambda_{(d)}^{7/9} M_{p}^{2/9}
\end{equation}
which, for $\Lambda_{(d)} = 10$ MeV, produces a bound $\Lambda_{(a)}<300$ GeV.  If instead, one requires $f_{th}$ to be only greater than the highest temperature achieved, which occurs when thermal friction gives way to mode instability -- when Eq.s \eqref{eq:steadyII} and \eqref{eq:steadyIII} are equal -- then one finds:
\begin{equation}
\Lambda_{(a)} < \Lambda_{(d)}^{17/27} M_{p}^{10/27}
\end{equation}
or for $\Lambda_{(d)} = 10$ MeV, $\Lambda_{(a)}<300$ TeV.

One can easily show that the lower stage field $\phi$, described in the previous section, does not evolve appreciably in its potential.  This in fact is not a phenomenological constraint, as the upper stage described in this section will always stop at a positive CC, allowing the lower stage to work (as long as $\phi$ is not at the bottom of its potential).  Nevertheless, one can check that during the upper stage, both classical rolling and evolution due to quantum fluctuations of $\phi$ are negligible.

In principle there could be several axion fields with varying scales in their potentials.  This could give rise to several stages of the relaxation model described here which could raise the initial CC far beyond the 300 GeV - 300 TeV scales found here.  We leave the exploration of multiple stages for future work.

\end{document}